# SOLUTION OF THE PROBLEM OF UNIQUENESS AND HERMITICITY OF HAMILTONIANS FOR DIRAC PARTICLES IN GRAVITATIONAL FIELDS

M.V. Gorbatenko, V.P. Neznamov

Russian Federal Nuclear Center – All-Russian Research Institute of Experimental Physics, Sarov, Mira 37, Nizhni Novgorod region, Russia, 607188
E-mail: neznamov@vniief.ru

**Abstract**

The authors prove that the dynamics of spin 1/2 particles in stationary gravitational fields can be described using an approach, which builds upon the formalism of pseudo-Hermitian Hamiltonians.

The proof consists in the analysis of three expressions for Hamiltonians, which are derived from the Dirac equation and describe the dynamics of spin 1/2 particles in the gravitational field of the Kerr solution. The Hamiltonians correspond to different choices of tetrad vectors and differ from each other. The differences between the Hamiltonians confirm the conclusion known from many studies that the Hamiltonians derived from the Dirac equation are non-unique. Application of standard pseudo-Hermitian quantum mechanics rules to each of these Hamiltonians produces the same Hermitian Hamiltonian.

The eigenvalue spectrum of the resulting Hamiltonian is the same as that of the Hamiltonians derived from the Dirac equation with any chosen system of tetrad vectors.

For description of the dynamics of spin 1/2 particles in stationary gravitational fields can be used not only the formalism of pseudo-Hermitian Hamiltonians, but also an alternative approach, which employs the Parker scalar product. The authors show that the alternative approach is equivalent to the formalism of pseudo-Hermitian Hamiltonians.

## 1. INTRODUCTION

In a number of studies (for example, in [1]), the attempts to describe the dynamics of half-integer spin particles in the gravitational field within the Hamiltonian formalism were found to encounter two problems that have not been resolved so far. One of the problems is that the form of resulting Hamiltonians depends on the choice of the system of tetrad vectors. A failure to resolve this problem will devalue the whole Hamiltonian formalism, because it is impossible to make physically verifiable predictions on its basis. The second problem is that Hamiltonians possess no Hermiticity, i.e. the key property underlying the apparatus of quantum mechanics and quantum field theory.

Authors of some papers (for example, [2], [3]) believe that the problem can be resolved by employing a relativistically invariant scalar product for state space vectors (the Parker scalar product). But this approach looks incomplete so far as in the publications there is no proof of Hermiticity of initial Hamiltonians at the use of the Parker scalar product.

In some papers (for example, in [4], [9]), the initial Hamiltonians were modified for the purpose of Hermiticity. This modification approach, however, did not seem substantiated enough.

In this paper, we propose to solve the problem of non-uniqueness and Hermiticity of Hamiltonians using the formalism of pseudo-Hermitian Hamiltonians (see [5]-[7] and references therein). The viability of this approach is supported by the following reasoning. Let us consider three types of Hamiltonians corresponding to a ½-spin particle in an axially symmetrical stationary gravitational field generated by a point body with mass $M$ and moment $\mathbf{J}$. Recall that the metric of such a gravitational field is described by the Kerr solution, which transforms into the Schwarzschild metric if $\mathbf{J} = 0$. Hamiltonians are written for different choices of the system of tetrad vectors:

1) in the Killing system of tetrad vectors [10], in which the tetrad vector $H_{\underline{0}}^{\alpha}$ coincides with the time passage direction of a distant observer;
2) in the system of tetrad vectors in the gauge of [11], [9];
3) in the system of tetrad vectors in the so-called symmetric gauge.

All the three Hamiltonians, first, differ from each other in their form, and, second, are not Hermitian. However, as shown in this paper, all the three Hamiltonians are pseudo-Hermitian.

Recall that in accordance with [5]-[7] the condition of pseudo-Hermiticity of the Hamiltonians assumes the existence of an invertible operator[1] $\rho$ satisfying the relationship

$$\rho \hat{H} \rho^{-1} = \hat{H}^{+}. \tag{1}$$

If there exists an operator $\eta$ satisfying the relationship

$$\rho = \eta \eta^{+}, \tag{2}$$

the Hamiltonian $\hat{\mathrm{H}}$ constructed in accordance with the law

$$\hat{\mathrm{H}} = \eta \hat{H} \eta^{-1} \tag{3}$$

will be Hermitian

$$\hat{\mathrm{H}} = \hat{\mathrm{H}}^{+} \tag{4}$$

---

[1] An operator $\rho$ in the papers [6] is called a metric operator. But we do not use this term with regard to $\rho$.

and will have an eigenvalue spectrum coinciding with that of the initial Hamiltonian $\hat{H}$. We will call the Hamiltonian $\hat{H}$ constructed by rule (3) the Hamiltonian in $\eta$-representation.

Note that not all of initially non-Hermitian Hamiltonians $\hat{H}$ are pseudo-Hermitian, and not all of them can be reduced to Hermitian expressions $\hat{H}$. However, for the three aforementioned types of Hamiltonians subject to the pseudo-Hermitian Hamiltonian formalism procedures, we draw conclusions, which are far from being evident a priory. For example, in all the three cases the expressions for $\hat{H}$ turn out, first, to be Hermitian, and, second, to be identical. Such an identity means that whatever the choice of the system of tetrad vectors in the gravitational field there will always exist a single Hermitian Hamiltonian $\hat{H}$, which has the same spectrum of energy levels as any of the starting operators $\hat{H}$. Upon transition to the Hamiltonian $\hat{H}$, one can use the quantum mechanics apparatus in its standard form. In particular, the left-hand side of the Schrödinger equation will contain the operator $i\hbar(\partial/\partial t)$, in which the time coordinate $t$ is understood to be the time of an infinitely distant observer.

As for the wave functions, in the formalism of pseudo-Hermitian Hamiltonians they will certainly change with the $\hat{H}$-to-$\hat{H}$ transition. The operator $\eta$ does not depend on time in the considered cases. So, if[2] $\Psi$ is a wave function in $\eta$-representation i.e. it satisfies the equation

$$i\frac{\partial \Psi}{\partial t} = \hat{H}\Psi, \qquad (5)$$

the appropriate wave function $\psi$ for the initial Hamiltonian i.e. the function satisfying to the equation

$$i\frac{\partial \psi}{\partial t} = \hat{H}\psi \qquad (6)$$

is calculated as

$$\Psi = \hat{\eta}\psi. \qquad (7)$$

The scalar products of the wave functions in $\eta$-representation have a standard form within the quantum mechanics[3] i.e.

$$(\Phi, \Psi) = \int d^3x \left(\Phi^+ \Psi\right). \qquad (8)$$

The scalar product for the wave functions in initial representation by definition is

$$\langle \varphi, \psi \rangle_\rho = \int d^3x \left(\varphi^+ \eta^+ \eta \psi\right) = \int d^3x \left(\varphi^+ \rho \psi\right). \qquad (9)$$

In this case

$$(\Phi, \Psi) = \langle \varphi, \psi \rangle_\rho. \qquad (10)$$

The Hermiticity condition for pseudo-Hermitian Hamiltonians is fulfilled concerning the scalar product (9) (see e.g. the relations (1)-(4))[4]:

---

[2] We use the capital Greek letters for the wave function in $\eta$-presentation and the small Greek letters for the wave function in initial presentation.

[3] The standard formalism of quantum mechanics we call a formalism in form which is stated in quantum mechanics textbooks and monographs (for example, [8]).

[4] We do not discuss here in detail the mathematical aspect of states spaces theory in pseudo-Hermitian quantum mechanics. The readers can find the theory, for example, in the paper [6].

$$\left\langle \varphi, \left(\hat{H}\psi\right)\right\rangle_\rho - \left\langle \left(\hat{H}\varphi\right), \psi\right\rangle_\rho =$$
$$= \int d^3x \left(\varphi^+ \rho \hat{H} \psi\right) - \int d^3x \left(\varphi^+ \hat{H}^+ \rho \psi\right) = 0. \quad (11)$$

We investigate the connection between formulas (9) for the scalar products and formation rules of scalar products of wave functions introduced in the papers [2], [3]. As a result we established that the operator $\rho$ in (9) coincides with the operator which follows from the rule [2], [3]. These results are also presented in this paper. In Section 7 we discuss the obtained results and come to the conclusion that in this paper the problem of uniqueness and Hermiticity of Hamiltonians for Dirac particles in gravitational fields is resolved.

## 2. THE REDUCING OF THE DIRAC EQUATION TO THE FORM OF THE SCHRÖDINGER EQUATION

The problem of spin 1/2 particle motion is usually (see e.g. - [12]-[15], [4], [9]) treated under two assumptions:
  (1) stationary of gravitational fields (Schwarzschild, Kerr),
  (2) minimality of interaction between bispinors and gravitational field, i.e. interaction appeared at writing of Dirac equation in the covariant form and accounting of dependence between global and local Dirac matrices by means of tetrad vectors $H_{\underline{\alpha}}^\mu$.

Recall the line of corresponding reasoning and introduce notations.

The tetrad vectors are defined by the relationships

$$H_{\underline{\alpha}}^\mu H_{\underline{\beta}}^\nu g_{\mu\nu} = \eta_{\underline{\alpha}\underline{\beta}}, \quad (12)$$

where

$$\eta_{\underline{\alpha}\underline{\beta}} = \text{diag}[-1,1,1,1]. \quad (13)$$

In addition to the system of tetrad vectors $H_{\underline{\alpha}}^\mu$, one can introduce three other systems of tetrad vectors, $H_{\underline{\alpha}\mu}$, $H^{\underline{\alpha}\mu}$, $H_\mu^{\underline{\alpha}}$, which differ from $H_{\underline{\alpha}}^\mu$ in the location of the global and local (underlined) indices. The global indices are raised up and lowered by means of the metric tensor $g_{\mu\nu}$ and inverse tensor $g^{\mu\nu}$, and the local indices are also raised up and lowered by means of the tensors $\eta_{\underline{\alpha}\underline{\beta}}$, $\eta^{\underline{\alpha}\underline{\beta}}$.

We assume that particle motion is described by the Dirac equation, which is written in the units $\hbar = c = 1$ as

$$\gamma^\alpha \left(\frac{\partial \psi}{\partial x^\alpha} + \Phi_\alpha \psi\right) - m\psi = 0. \quad (14)$$

Here, $m$ is the particle mass, $\psi$ is the four-component "column" bispinor, and $\gamma^\alpha$ are the $4 \times 4$ Dirac matrices satisfying the relationship

$$\gamma^\alpha \gamma^\beta + \gamma^\beta \gamma^\alpha = 2g^{\alpha\beta} E. \quad (15)$$

Here $E$ is $4 \times 4$ identity matrix.

The parentheses in (14) contains a covariant derivative of the bispinor $\nabla_\alpha \psi$,

$$\nabla_\alpha \psi = \frac{\partial \psi}{\partial x^\alpha} + \Phi_\alpha \psi. \quad (16)$$

The structure $\nabla_\alpha \psi$ contains the quantity $\Phi_\alpha$ called bispinor connectivity. To find the explicit form of $\Phi_\alpha$, one should fix some system of tetrad vectors $H_{\underline{\alpha}}^\mu$ defined by the relationship (12). Upon that, the quantity $\Phi_\alpha$ can be expressed through Christoffel derivatives from tetrad in the following way (the Christoffel derivatives are denoted by the semicolon):

$$\Phi_\alpha = -\frac{1}{4} H_\mu^{\underline{\varepsilon}} H_{\nu\underline{\varepsilon};\alpha} S^{\mu\nu}. \quad (17)$$

In what follows, along with Dirac matrices with global indices $\gamma^\alpha$, we will use Dirac matrices with local indices $\gamma^{\underline{\alpha}}$. The relationship between $\gamma^\alpha$ and $\gamma^{\underline{\alpha}}$ is given by the expression

$$\gamma^\alpha = H_{\underline{\beta}}^\alpha \gamma^{\underline{\beta}}. \quad (18)$$

It follows from (18), (15), (12), that

$$\gamma^{\underline{\alpha}} \gamma^{\underline{\beta}} + \gamma^{\underline{\beta}} \gamma^{\underline{\alpha}} = 2\eta^{\underline{\alpha\beta}} E. \quad (19)$$

In terms of the matrices $\gamma^{\underline{\alpha}}$, the Dirac equation (14) can be written as follows:

$$H_{\underline{\mu}}^\alpha \gamma^{\underline{\mu}} \left( \frac{\partial \psi}{\partial x^\alpha} + \Phi_\alpha \psi \right) - m\psi = 0. \quad (20)$$

It is convenient to choose the quantities $\gamma^{\underline{\alpha}}$ so that they have the same form for all local frames of reference. The sets $\gamma^{\underline{\alpha}}$ and $\gamma^\alpha$ can be used to construct a full set of $4 \times 4$ matrices. The full set is for example the set

$$E, \quad \gamma_\alpha, \quad S_{\alpha\beta} \equiv \frac{1}{2}(\gamma_\alpha \gamma_\beta - \gamma_\beta \gamma_\alpha), \quad \gamma_5 \equiv \gamma_0 \gamma_1 \gamma_2 \gamma_3, \quad \gamma_5 \gamma_\alpha. \quad (21)$$

Any set of Dirac matrices provides for several discrete automorphisms. We restrict ourselves to the automorphism

$$\gamma_\alpha \to \gamma_\alpha^+ = -D\gamma_\alpha D^{-1}. \quad (22)$$

The matrix $D$ will be called anti-hermitizing.

The Dirac equation can be written in the form of the Schrödinger equation. If the resulting Hamiltonian operator (or the evolution operator) were Hermitian and unique, it would be easier to analyze the theory's physical content, because this would allow us to employ the apparatus of the quantum theory for finding the spectrum of eigenvalues and state vectors. In the general case, however, the Hamiltonian operator does not possess the aforementioned properties. Let us consider this issue in more detail.

The way of reducing the Dirac equation (14) to the form of the Schrödinger equation is determined by the requirement that the left-hand side of the Schrödinger equation should contain the operator $i(\partial/\partial t)$, with $t$ being the time of a distant observer.

$$i \frac{\partial \Psi}{\partial t} = \hat{H} \Psi. \quad (23)$$

The operator $\hat{H}$ on the right side of (23) has the meaning of an evolution operator in the frame of reference related to the distant observer. It follows from (14), (23), that

$$\hat{H} = \frac{i}{(-g^{00})} \gamma^0 \gamma^k \nabla_k - i\Phi_0 - \frac{im}{(-g^{00})} \gamma^0. \quad (24)$$

By substituting expression (16) for covariant derivatives in (24), we obtain:

$$\hat{H} = -\frac{im}{(-g^{00})}\gamma^0 + \frac{i}{(-g^{00})}\gamma^0\gamma^k\frac{\partial}{\partial x^k} - i\Phi_0 + \frac{i}{(-g^{00})}\gamma^0\gamma^k\Phi_k. \qquad (25)$$

In order to further refine the expression for $\hat{H}$, one should substitute expressions (17) for $\Phi_0$ and $\Phi_k$ in (25). As (17) contains tetrad vectors and their derivatives, it turns out after this kind of substitution that the Hamiltonian becomes dependent on the choice of tetrad vectors. This fact might seem non-physical, if we take into account that the Dirac equation in the initial form is invariant with respect to the transition from one system of tetrad vectors to another. Writing the Dirac equation in the Schrödinger form (23), however, assumes the diversion from covariance, because the partial time derivative is transferred to the left side, which distinguishes the time coordinate. This very coordinate disparity results in the Hamiltonian's sensitivity to the choice of the system of tetrad vectors. Another "drawback" of the Hamiltonian written as (25) is the lack of Hermiticity in standard understanding of this term (see Introduction).

The above considerations about uniqueness and Hermiticity of Hamiltonian (25) are the problems which are solved in the further sections.

## 3. THE KERR SOLUTION FOR A WEAK GRAVITATIONAL FIELD

The metric of the Kerr solution in the first-order approximation in "mass" $M$ with length dimensionality ($M$ is half the gravitational radius, i.e. the quantity $GM/c^2$) is presented in many references, for example, in [16], [17]. In this approximation, the Kerr metric has the following form:

$$\left. \begin{array}{l} g_{00} = -1 + 2\dfrac{M}{R}; \quad g_{0k} = 2\dfrac{M(J_{kl}R_l)}{R^3}; \quad g_{mn} = \delta_{mn} + 2\dfrac{M}{R}\delta_{mn}; \\[4pt] \sqrt{-g} = 1 + 2\dfrac{M}{R}; \\[4pt] g^{00} = -1 - 2\dfrac{M}{R}; \quad g^{0k} = 2\dfrac{M(J_{kl}R_l)}{R^3}; \quad g^{mn} = \delta_{mn} - 2\dfrac{M}{R}\delta_{mn}. \end{array} \right\} \qquad (26)$$

Let us call the quantity $J_{mn}$ in (26) the "reduced tensor of angular momentum" (with $J_{mn}$ being the quantity derived by dividing the "physical" tensor of angular momentum by $Mc$). Instead of the tensor $J_{mn}$, one can use its equivalent, the axial tensor $J_k = \tfrac{1}{2}\varepsilon_{mnk}J_{mn}$, where $\varepsilon_{mnk}$ is a totally anti-symmetric third-rank tensor.

The Christoffel symbols corresponding to the metric (26) are equal to:

$$\left. \begin{array}{l} \begin{pmatrix} 0 \\ 00 \end{pmatrix} = 0; \quad \begin{pmatrix} 0 \\ 0k \end{pmatrix} = \dfrac{MR_k}{R^3}; \quad \begin{pmatrix} 0 \\ mn \end{pmatrix} = 3\dfrac{M\left[(J_{ml}R_l)R_n + (J_{nl}R_l)R_m\right]}{R^5}; \\[6pt] \begin{pmatrix} k \\ 00 \end{pmatrix} = \dfrac{MR_k}{R^3}; \quad \begin{pmatrix} m \\ 0n \end{pmatrix} = 2\dfrac{MJ_{mn}}{R^3} - 3\dfrac{M\left[(J_{ml}R_l)R_n - (J_{nl}R_l)R_m\right]}{R^5}; \\[6pt] \begin{pmatrix} k \\ mn \end{pmatrix} = -\dfrac{M}{R^3}\left[\delta_{km}R_n + \delta_{kn}R_m - \delta_{mn}R_k\right]. \end{array} \right\} \qquad (27)$$

In many publications (for example in [4]), the dynamics of a half-integer spin particle was studied in the space, the metric of which was described by two functions of spatial coordinates: $V = V(x,y,z)$, $W = W(x,y,z)$.

$$ds^2 = -V^2 dt^2 + W^2 \left(dx^2 + dy^2 + dz^2\right). \tag{28}$$

By comparing (28) with (26), we see that for the Schwarzschild solution $(J_{kl} = 0)$

$$V = 1 - \frac{M}{R}; \quad W = 1 + \frac{M}{R}. \tag{29}$$

Note that expression (28) for the square of interval has a limited range of applicability. For example, the Kerr solution (26) in principle cannot be reduced to (28). In what follows, we will use the notations of (26).

For the Kerr field, let us introduce a Killing vector, which is associated with the time direction as viewed by a distant observer. There are three Killing vectors in the Schwarzschild gravitational field, and two in the Kerr field. The Killing vectors are defined through the Lie derivative. The Lie derivative of the metric along the vector $\xi^\alpha$ has the following form:

$$\delta_\xi g_{\alpha\beta} = \xi^\sigma g_{\alpha\beta,\sigma} + \xi^\sigma{}_{,\alpha} g_{\sigma\beta} + \xi^\sigma{}_{,\beta} g_{\alpha\sigma}. \tag{30}$$

If the vector $\xi^\alpha$ is such that the equality

$$\delta_\xi g_{\alpha\beta} = 0 \tag{31}$$

is fulfilled, the vector $\xi^\alpha$ is a Killing vector. It is easy to see that the vector of the form

$$\xi^\alpha = (1,0,0,0) \tag{32}$$

is a Killing vector in case of stationary metric.

By definition, the Killing system of tetrad vectors will be called the system, in which the tetrad vector $\tilde{H}^\alpha_{\underline{0}}$ is collinear with the vector $\xi^\alpha$, i.e. the tetrad vector has only non-zero component $\tilde{H}^0_{\underline{0}}$.

$$\tilde{H}^0_{\underline{0}} \neq 0, \quad \tilde{H}^k_{\underline{0}} = 0. \tag{33}$$

The vectors that satisfy (33) will be denoted with an upper tilde.

Using (12) and (33), we establish that[5]:

$$\tilde{H}^0_{\underline{0}} = \left[1 + \frac{M}{R}\right]; \quad \tilde{H}^k_{\underline{0}} = 0; \quad \tilde{H}^0_{\underline{k}} = 2\frac{M(J_{kl}R_l)}{R^3}; \quad \tilde{H}^m_{\underline{k}} = \delta_{mk}\left[1 - \frac{M}{R}\right]. \tag{34}$$

Note that expressions (34) are the same as those obtained for a stationary gravitational field in [10]. However, they differ from the expressions in [11], which were used in some papers (for example, in [9]).

The quantities $\Phi_\alpha$ are found using formulas (17). For equations (34) we obtain:

$$\tilde{\Phi}_0 = -\frac{1}{2}\frac{MR_k}{R^3}S^{0k} + \frac{1}{2}\left\{\frac{MJ_{mn}}{R^3} - \frac{3}{2}\frac{M\left[(J_{ml}R_l)R_n - (J_{nl}R_l)R_m\right]}{R^5}\right\}S^{mn}; \tag{35}$$

---

[5] Here and further we give nothing but one for each four possible systems of tetrad vectors. The other three systems are obtained by means of rising and lowering of global indices with the help of the metric tensor $g_{\mu\nu}$ and the inverse tensor $g^{\mu\nu}$ and for the local indices with the help of tensors $\eta_{\underline{\alpha\beta}}$, $\eta^{\underline{\alpha\beta}}$ (see explanation for (12)).

$$\tilde{\Phi}_k = \frac{1}{4}\frac{M}{R^3}\left[R_p\delta_{kq} - R_q\delta_{kp}\right]S^{\underline{pq}} - \left\{\frac{MJ_{km}}{R^3} - \frac{3}{2}\frac{M\left[(J_{kl}R_l)R_m - (J_{ml}R_l)R_k\right]}{R^5}\right\}\gamma_{\underline{0}}\gamma^{\underline{m}}. \quad (36)$$

## 4. Hamiltonians for the Kerr solution

### 4.1. Hamiltonian in the Killing system of tetrad vectors

The Hamiltonian operator $\hat{\tilde{H}}$ is derived from general formula (25), but using tetrad vectors with tildes (34) and bispinor connectivity components with tildes (35), (36).

$$\hat{\tilde{H}} = -\frac{im}{(-g^{00})}\tilde{\gamma}^{\underline{0}} + \frac{i}{(-g^{00})}\tilde{\gamma}^{\underline{0}}\tilde{\gamma}^{\underline{k}}\frac{\partial}{\partial x^k} - i\tilde{\Phi}_0 + \frac{i}{(-g^{00})}\tilde{\gamma}^{\underline{0}}\tilde{\gamma}^{\underline{k}}\tilde{\Phi}_k. \quad (37)$$

For convenience of calculations, Hamiltonian (37) is written as a sum of four summands.

$$\hat{\tilde{H}} \equiv \hat{\tilde{H}}_1 + \hat{\tilde{H}}_2 + \hat{\tilde{H}}_3 + \hat{\tilde{H}}_4. \quad (38)$$

Here:

$$\hat{\tilde{H}}_1 \equiv -\frac{im}{(-g^{00})}\tilde{\gamma}^{\underline{0}}, \quad (39)$$

$$\hat{\tilde{H}}_2 \equiv \frac{i}{(-g^{00})}\tilde{\gamma}^{\underline{0}}\tilde{\gamma}^{\underline{k}}\frac{\partial}{\partial x^k}, \quad (40)$$

$$\hat{\tilde{H}}_3 \equiv -i\tilde{\Phi}_0, \quad (41)$$

$$\hat{\tilde{H}}_4 \equiv \frac{i}{(-g^{00})}\tilde{\gamma}^{\underline{0}}\tilde{\gamma}^{\underline{k}}\tilde{\Phi}_k. \quad (42)$$

We calculate each of the four summands in expression (38).

$$\hat{\tilde{H}}_1 = im\left(1 - \frac{M}{R}\right)\gamma_{\underline{0}} - 2im\frac{M(J_{kl}R_l)}{R^3}\gamma^{\underline{k}}; \quad (43)$$

$$\hat{\tilde{H}}_2 = -i\left[1 - 2\frac{M}{R}\right]\gamma_{\underline{0}}\gamma^{\underline{k}}\frac{\partial}{\partial x^k} + 2i\frac{M(J_{kl}R_l)}{R^3}\frac{\partial}{\partial x^k} + 2i\frac{M(J_{ml}R_l)}{R^3}S_{\underline{mk}}\frac{\partial}{\partial x^k}; \quad (44)$$

$$\hat{\tilde{H}}_3 = -\frac{i}{2}\frac{MR_k}{R^3}\gamma_{\underline{0}}\gamma_{\underline{k}} + \frac{i}{2}\left\{\frac{M}{R^3}J_k - 3\frac{M(J_lR_l)R_k}{R^5}\right\}\gamma_{\underline{5}}\gamma_{\underline{0}}\gamma_{\underline{k}}; \quad (45)$$

$$\hat{\tilde{H}}_4 = i\frac{M}{R^3}R_k\gamma_{\underline{0}}\gamma^{\underline{k}} - i\frac{MJ_k}{R^3}\gamma_{\underline{5}}\gamma_{\underline{0}}\gamma_{\underline{k}} + 3i\frac{M(J_lR_l)R_k}{R^5}\gamma_{\underline{5}}\gamma_{\underline{0}}\gamma_{\underline{k}}. \quad (46)$$

By substituting (43)-(46) in (38) we obtain:

$$\hat{\tilde{H}} = im\gamma_{\underline{0}} - im\frac{M}{R}\gamma_{\underline{0}} - i\gamma_{\underline{0}}\gamma^{\underline{k}}\frac{\partial}{\partial x^k} + 2i\frac{M}{R}\gamma_{\underline{0}}\gamma^{\underline{k}}\frac{\partial}{\partial x^k} + \frac{i}{2}\frac{MR_k}{R^3}\gamma_{\underline{0}}\gamma^{\underline{k}} +$$
$$+2i\frac{M(J_{kl}R_l)}{R^3}\frac{\partial}{\partial x^k} -$$
$$-2im\frac{M(J_{kl}R_l)}{R^3}\gamma^{\underline{k}} + 2i\frac{M(J_{ml}R_l)}{R^3}S_{\underline{mk}}\frac{\partial}{\partial x^k} -$$
$$-\frac{i}{2}\left\{\frac{M}{R^3}J_k - 3\frac{M(J_l R_l)R_k}{R^5}\right\}\gamma_{\underline{5}}\gamma_{\underline{0}}\gamma_{\underline{k}}. \qquad (47)$$

The terms that do not contain $J_{mn}$ correspond to the Hamiltonian in the Schwarzschild problem. They are the same as the corresponding expression for the Hamiltonian in [4]. The rest of Hamiltonian (47), i.e. the part with $J_{mn}$, however, differs from a similar part in [9]. This stems from the difference in the system of tetrad vectors. This fact is also evidenced by considerations presented in [1], which points out that the "ambiguity" of Hamiltonians is due to their dependence on the choice of tetrad vectors.

The tensor $J_{mn}$ in expression (47) can be replaced with an axial tensor $J_k$. Such a replacement will result in the following form of the Hamiltonian $\hat{\tilde{H}}$ in the Killing system of reference vectors:

$$\hat{\tilde{H}} = im\gamma_{\underline{0}} - im\frac{M}{R}\gamma_{\underline{0}} - i\gamma_{\underline{0}}\gamma^{\underline{k}}\frac{\partial}{\partial x^k} + 2i\frac{M}{R}\gamma_{\underline{0}}\gamma^{\underline{k}}\frac{\partial}{\partial x^k} + \frac{i}{2}\frac{MR_k}{R^3}\gamma_{\underline{0}}\gamma^{\underline{k}} +$$
$$+2i\frac{M}{R^3}\varepsilon_{klq}R_l J_q\frac{\partial}{\partial x^k} - 2im\frac{M}{R^3}\varepsilon_{klq}R_l J_q\gamma_{\underline{k}} + 2i\frac{M}{R^3}(J_l\gamma_{\underline{5}}\gamma_{\underline{0}}\gamma_{\underline{l}})R_k\frac{\partial}{\partial x^k} -$$
$$-2i\frac{M}{R^3}(R_l\gamma_{\underline{5}}\gamma_{\underline{0}}\gamma_{\underline{l}})J_k\frac{\partial}{\partial x^k} - \frac{i}{2}\left\{\frac{M}{R^3}J_k - 3\frac{M(J_l R_l)R_k}{R^5}\right\}\gamma_{\underline{5}}\gamma_{\underline{0}}\gamma_{\underline{k}}. \qquad (48)$$

Obviously expressions (47) and (48) are equivalent.

### 4.2. Two other Hamiltonians

In addition to Hamiltonian (48), let us present two other expressions for Hamiltonians in another systems of tetrad vectors without their derivation using the notations adopted above.

1) System of tetrad vectors used in [11], [9]:

$$H_{\underline{0}}^{\prime 0} = \left[1 + \frac{M}{R}\right]; \quad H_{\underline{0}}^{\prime k} = -2\frac{M(J_{kl}R_l)}{R^3}; \quad H_{\underline{k}}^{\prime 0} = 0; \quad H_{\underline{k}}^{\prime m} = \delta_{mk}\left[1 - \frac{M}{R}\right]. \qquad (49)$$

Hamiltonian in system of tetrad vectors (49):

$$\hat{H}' = im\gamma_{\underline{0}} - im\frac{M}{R}\gamma_{\underline{0}} - i\gamma_{\underline{0}}\gamma_{\underline{k}}\frac{\partial}{\partial x^k} + 2i\frac{M}{R}\gamma_{\underline{0}}\gamma^{\underline{k}}\frac{\partial}{\partial x^k} + \frac{i}{2}\frac{MR_k}{R^3}\gamma_{\underline{0}}\gamma^{\underline{k}} +$$
$$+2i\frac{M(J_{kl}R_l)}{R^3}\frac{\partial}{\partial x^k} +$$
$$+\frac{i}{2}\left\{\frac{M}{R^3}J_k - 3\frac{M(J_l R_l)R_k}{R^5}\right\}\gamma_{\underline{5}}\gamma_{\underline{0}}\gamma_{\underline{k}}. \qquad (50)$$

2) System of tetrad vectors in symmetric gauge:

$$H_{\underline{0}}^0 = \left[1 + \frac{M}{R}\right]; \quad H_{\underline{0}}^k = -\frac{M(J_{kl}R_l)}{R^3}; \quad H_{\underline{k}}^0 = \frac{M(J_{kl}R_l)}{R^3}; \quad H_{\underline{k}}^m = \delta_{mk}\left[1 - \frac{M}{R}\right]. \quad (51)$$

Hamiltonian in system of tetrad vectors (51):

$$\begin{aligned}
\hat{H} &= im\gamma_{\underline{0}} - i\gamma_{\underline{0}}\gamma_{\underline{k}}\frac{\partial}{\partial x^k} - im\frac{M}{R}\gamma_{\underline{0}} + 2i\frac{M}{R}\gamma_{\underline{0}}\gamma^{\underline{k}}\frac{\partial}{\partial x^k} + \frac{i}{2}\frac{MR_k}{R^3}\gamma_{\underline{0}}\gamma^{\underline{k}} + \\
&\quad + 2i\frac{M(J_{kl}R_l)}{R^3}\frac{\partial}{\partial x^k} - \\
&\quad - im\frac{M(J_{kl}R_l)}{R^3}\cdot\gamma_{\underline{k}} + i\frac{M(J_{ml}R_l)}{R^3}S_{\underline{mk}}\frac{\partial}{\partial x^k}.
\end{aligned} \quad (52)$$

In the Hamiltonians, (47), (50), (52), the Schwarzschild terms and one of the Kerr terms are the same. The rest of the terms are different for all the three Hamiltonians. The widest range of terms is observed in the Hamiltonian $\hat{\tilde{H}}$.

None of the Hamiltonians, (47), (50), (52), is Hermitian.

### 4.3. HAMILTONIANS IN THE NOTATIONS OF [4], [9]

For convenience of comparison with results of [4], [9], the expressions (48), (50), (52) are presented for the case of the metric $\eta_{\underline{\alpha\beta}} = \text{diag}[1,-1,-1,-1]$ and Dirac matrices $\beta = \gamma^0$, $\alpha^i = \gamma^0\gamma^i$, $\Sigma^k = \gamma_5\gamma^0\gamma^k$. Expressions (48), (50), (52) take the form of expressions (53), (54), (55) respectively[6].

$$\begin{aligned}
\hat{\tilde{H}} &= \beta m - \frac{M}{R}\beta m + \boldsymbol{\alpha}\mathbf{p} - \frac{2M}{R}\boldsymbol{\alpha}\mathbf{p} + \frac{i}{2}\frac{M}{R^3}\boldsymbol{\alpha}\mathbf{R} + \\
&\quad + \frac{2M}{R^3}\mathbf{J}(\mathbf{R}\times\mathbf{p}) + \\
&\quad + \frac{2M}{R^3}(\mathbf{J}\times\mathbf{R})\beta m\boldsymbol{\alpha} - \frac{2M}{R^3}i(\boldsymbol{\Sigma}\mathbf{J})(\mathbf{R}\mathbf{p}) + \frac{2M}{R^3}i(\boldsymbol{\Sigma}\mathbf{R})(\mathbf{J}\mathbf{p}) + \\
&\quad + \frac{1}{2}\left(\frac{M}{R^3}(\boldsymbol{\Sigma}\mathbf{J}) - \frac{3M(\mathbf{J}\mathbf{R})(\boldsymbol{\Sigma}\mathbf{R})}{R^5}\right).
\end{aligned} \quad (53)$$

$$\begin{aligned}
\hat{H}' &= \beta m - \frac{M}{R}\beta m + \boldsymbol{\alpha}\mathbf{p} - \frac{2M}{R}\boldsymbol{\alpha}\mathbf{p} + \frac{i}{2}\frac{M}{R^3}\boldsymbol{\alpha}\mathbf{R} + \\
&\quad + \frac{2M}{R^3}\mathbf{J}(\mathbf{R}\times\mathbf{p}) - \\
&\quad - \frac{1}{2}\left(\frac{M}{R^3}(\boldsymbol{\Sigma}\mathbf{J}) - \frac{3M(\mathbf{J}\mathbf{R})(\boldsymbol{\Sigma}\mathbf{R})}{R^5}\right).
\end{aligned} \quad (54)$$

---

[6] In (50), (52) tensors $J_{mn}$ substitute for axial vectors $J_k$.

$$\hat{H} = \beta m - \frac{M}{R}\beta m + \boldsymbol{\alpha}\mathbf{p} - \frac{2M}{R}\boldsymbol{\alpha}\mathbf{p} + \frac{i}{2}\frac{M}{R^3}\boldsymbol{\alpha}\mathbf{R} +$$
$$+ \frac{2M}{R^3}\mathbf{J}(\mathbf{R}\times\mathbf{p}) - \quad (55)$$
$$- \frac{M}{R^3}(\mathbf{J}\times\mathbf{R})\beta m\boldsymbol{\alpha} - \frac{M}{R^3}i(\boldsymbol{\Sigma}\mathbf{J})(\mathbf{R}\mathbf{p}) + \frac{M}{R^3}i(\boldsymbol{\Sigma}\mathbf{R})(\mathbf{J}\mathbf{p}).$$

## 5. FORMALISM OF PSEUDO-HERMITIAN HAMILTONIANS

### 5.1. CONSTRUCTION OF THE OPERATOR $\hat{H}$ IN SYSTEM OF TETRAD VECTORS [11], [9]

First, we need to make sure that Hamiltonian (50) is pseudo-Hermitian. The pseudo-Hermiticity condition (1) assumes the existence of operators $\rho$, $\rho^{-1}$ that satisfy the relationship $\rho\hat{H}\rho^{-1} = \hat{H}^\dagger$. If we write this relationship for Hamiltonian (50), to within the first order in $M$ we will obtain:

$$\rho = 1 + \frac{3M}{R}, \quad \rho^{-1} = 1 - 3\frac{M}{R}. \quad (56)$$

From (2) and (56) we obtain that

$$\eta = 1 + \frac{3}{2}\frac{M}{R}. \quad (57)$$

By substituting (57) in (3) and using expression (50), we find the desired expression for the Hermitian Hamiltonian:

$$\hat{H} = \left(1 + \frac{3}{2}\frac{M}{R}\right)\hat{H}\left(1 - \frac{3}{2}\frac{M}{R}\right) =$$
$$= im\gamma_0 - im\frac{M}{R}\gamma_0 - i\gamma_0\gamma_k\frac{\partial}{\partial x^k} + 2i\frac{M}{R}\gamma_0\gamma^k\frac{\partial}{\partial x^k} - i\frac{MR_k}{R^3}\gamma_0\gamma^k +$$
$$+ 2i\frac{M(J_{kl}R_l)}{R^3}\frac{\partial}{\partial x^k} + \quad (58)$$
$$+ \frac{i}{2}\left\{\frac{M}{R^3}J_k - 3\frac{M(J_l R_l)R_k}{R^5}\right\}\gamma_5\gamma_0\gamma_k.$$

By comparing (58) with (50), we see that only the Schwarzschild part has changed. The Kerr term in Hamiltonian (50) was Hermitian and it has not been affected by the transformation of (3).
Expression (58) is completely the same as the Hermitian Hamiltonian used in [9].

### 5.2. CONSTRUCTION OF THE OPERATOR $\hat{H}$ FOR OTHER SYSTEMS OF TETRAD VECTORS

Hamiltonians (47), (52), similar to Hamiltonian (50), are pseudo-Hermitian.

The procedure of deriving the Hermitian operator $\hat{H}$ for Hamiltonians (47), (52) is identical to the above procedure for Hamiltonian (50) and will not be described here.

As a result of implementing this procedure as applied to Hamiltonian (47), the operator $\eta$ for it was found, first, to be equal to

$$\eta = 1 + \frac{3}{2}\frac{M}{R} + \frac{M(J_{km}R_m)}{R^3}\gamma_0\gamma_k. \tag{59}$$

Second, the resulting Hermitian Hamiltonian is the same as Hamiltonian (58).

As a result of implementing this procedure as applied to Hamiltonian (52), the operator $\eta$ for it was found, first, to be equal to

$$\eta = 1 + \frac{3}{2}\frac{M}{R} + \frac{1}{2}\frac{M(J_{km}R_m)}{R^3}\gamma_0\gamma_k. \tag{60}$$

Second, the resulting Hermitian Hamiltonian is also the same as Hamiltonian (58).

### 5.3. SUMMARY OF PSEUDO-HERMITIAN HAMILTONIANS

The approach employing the formalism of pseudo-Hermitian Hamiltonians leads to the standard apparatus of quantum mechanics. The resulting Hamiltonian generated within this approach is expressed as (58). It
  - does not depend on the choice of tetrad vectors,
  - is Hermitian,
  - has an eigenvalue spectrum, which coincides with that of initial Hamiltonians for any choice of tetrad vectors.

The scalar products can be calculated by two ways. The scalar products are calculated by equation (8) in terms of the wave functions in $\eta$ - representation. For calculation of scalar products in terms of the wave functions in initial representation it is necessary to use the equations (9) which content the invertible operators $\rho$ explicitly. For convenience, the quantities $\rho$ and $\eta$ for three systems of tetrad vectors are presented in Table 1. It should be remembered that in all the three systems of tetrad vectors, the Hamiltonian $\hat{H}$ in $\eta$-representation is Hermitian and has the same form and is defined as (58).

Table 1 – Operators $\rho$ and $\eta$ for three systems of tetrad vectors

| Tetrad vectors | Operator $\rho$ | Operator $\eta$ |
|---|---|---|
| Killing system (34) | $1 + 3\frac{M}{R} + 2\frac{M(J_{km}R_m)}{R^3}\gamma_0\gamma_k$ | $1 + \frac{3}{2}\frac{M}{R} + \frac{M(J_{km}R_m)}{R^3}\gamma_0\gamma_k$ |
| System (49) – system of tetrad vectors in gauge [11], [9] | $1 + 3\frac{M}{R}$ | $1 + \frac{3}{2}\frac{M}{R}$ |
| System (51) – system of tetrad vectors in symmetric gauge | $1 + 3\frac{M}{R} + \frac{M(J_{km}R_m)}{R^3}\gamma_0\gamma_k$ | $1 + \frac{3}{2}\frac{M}{R} + \frac{1}{2}\frac{M(J_{km}R_m)}{R^3}\gamma_0\gamma_k$ |

# 6. THE PARKER SCALAR PRODUCT

## 6.1. THE SCALAR PRODUCT INTRODUCED IN [2], [3]

Let us consider of approach based on use of the scalar product of wave functions introduced in [2], [3] (let us call this scalar product the Parker scalar product). Let us provide some auxiliary relationships, which are convenient as applied to the Parker formalism.

It follows from (18), (22) that the operator of anti-Hermitian conjugation of matrices $\gamma_\alpha$ with global indices is the same matrix $\gamma_{\underline{0}}$, which performs these functions for the matrices $\gamma_{\underline{\alpha}}$ with local indices. Thus,

$$\gamma_{\underline{\alpha}}^+ = \gamma_{\underline{0}} \gamma_{\underline{\alpha}} \gamma_{\underline{0}}, \quad \gamma_\alpha^+ = \gamma_{\underline{0}} \gamma_\alpha \gamma_{\underline{0}}. \tag{61}$$

Using (61), we obtain:

$$\left(S_{\underline{\alpha\beta}}\right)^+ = \gamma_{\underline{0}} S_{\underline{\alpha\beta}} \gamma_{\underline{0}}, \quad \left(S_{\alpha\beta}\right)^+ = \gamma_{\underline{0}} S_{\alpha\beta} \gamma_{\underline{0}}, \quad \gamma_5^+ = -\gamma_{\underline{0}} \gamma_5 \gamma_{\underline{0}}. \tag{62}$$

Considering (62), and the fact that bispinor connectivity $\Phi_\alpha$ is expressed as

$$\Phi_\alpha = \frac{1}{4} \Phi_{\alpha\underline{\mu\nu}} S^{\underline{\mu\nu}}, \tag{63}$$

without loss of generality, we can write that

$$\left(\Phi_\alpha\right)^+ = \gamma_{\underline{0}} \Phi_\alpha \gamma_{\underline{0}} \quad \to \quad \Phi_\alpha = \gamma_{\underline{0}} \left(\Phi_\alpha\right)^+ \gamma_{\underline{0}}. \tag{64}$$

Covariant derivatives of Dirac matrices are equal to zero,

$$\nabla_\mu \gamma_\alpha = \gamma_{\alpha;\mu} + \left[\Phi_\mu, \gamma_\alpha\right]_- = 0. \tag{65}$$

Papers [2], [3] introduce the following rule for calculating the scalar product of two wave functions $\varphi$ and $\psi$:

$$\langle \varphi, \psi \rangle = \int dx^3 \sqrt{-g} \left(\varphi^+ \gamma_{\underline{0}} \gamma^0 \psi\right). \tag{66}$$

It is easy to see that the right-hand quantity is not only invariant with respect to the choice of tetrad vectors, but it is also a four-dimensional scalar with respect to the transformation of global coordinates. Indeed, for any two wave functions $\varphi$ and $\psi$, one can introduce a four-dimensional vector

$$j^\alpha \equiv \left(\varphi^+ \gamma_{\underline{0}} \gamma^\alpha \psi\right). \tag{67}$$

We choose a four-dimensional volume, $V_4$, and perform integration of the scalar $j^\alpha{}_{;\alpha}$ over it. We obtain a scalar $\int_{V_4} d^4x \sqrt{-g} \, j^\alpha{}_{;\alpha}$. Note that neither the scalar $j^\alpha{}_{;\alpha}$, nor the scalar obtained by integration is generally speaking equal to zero. The integral scalar can be transformed according to the Gauss theorem. If the four-dimensional space is chosen such that the three-dimensional spaces bounding it are orthogonal to the Killing vector, it will follow from the Gauss theorem that

$$\int_{V_4} d^4x \sqrt{-g} \, j^\alpha{}_{;\alpha} = \int_{V_3} d^3x \sqrt{-g} \, j^0. \tag{68}$$

Since the left side contains a scalar, the quantity $(\varphi, \psi)$ on the right side should also be a global scalar.

Note that in case of plane space, tetrad vectors can be identified with tangent vectors to the coordinate lines. In this case, the expression for the scalar product (67) in Cartesian coordinates will take the standard form for quantum mechanics:

$$\langle \varphi, \psi \rangle \to \int dx^3 \left( \varphi^+ \psi \right).$$

This property of the quantity $\langle \varphi, \psi \rangle$ can be treated as a manifestation of the correspondence principle.

So, we have to use the equation (66) for calculation the Parker scalar product $\langle \varphi, \psi \rangle$. Let us write this equation for the case using of Killing system of tetrad vectors (34).

$$\langle \varphi, \psi \rangle = \int dx^3 \sqrt{-g} \left( \varphi^+ \gamma_{\underline{0}} \gamma^0 \psi \right) =$$
$$= \int dx^3 \sqrt{-g} \left\{ \left( \varphi^+ \gamma_{\underline{0}} H^0_{\underline{0}} \gamma^0 \psi \right) + \left( \varphi^+ \gamma_{\underline{0}} H^0_{\underline{k}} \gamma^k \psi \right) \right\} = \quad (69)$$
$$= \int dx^3 \left( 1 + 2\frac{M}{R} \right) \sqrt{-g} \left\{ \left( \varphi^+ \left( 1 + 2\frac{M}{R} \right) \psi \right) + \left( \varphi^+ \gamma_{\underline{0}} \left( 2\frac{M(J_{kl} R_l)}{R^3} \right) \gamma^k \psi \right) \right\}.$$

After transformation we obtain

$$\langle \varphi, \psi \rangle = \int dx^3 \left( \varphi^+ \left( 1 + 3\frac{M}{R} + 2\frac{M(J_{kl} R_l)}{R^3} \gamma_{\underline{0}} \gamma^{\underline{k}} \right) \psi \right). \quad (70)$$

By comparing (70) with (9), we see that for the Killing system

$$\rho = 1 + 3\frac{M}{R} + 2\frac{M(J_{km} R_m)}{R^3} \gamma_{\underline{0}} \gamma_k, \quad (71)$$

i.e. the same value which was obtained in pseudo-Hermitian formalism (see Table 1). The similar results are obtained for two other system of tetrad vectors considered above. We omit the proof.

The obtained results mean that in case of stationary gravitational fields the use of the Parker scalar product results in practically the use of the invertible operator $\rho$ which follows from the formalism of pseudo-Hermitian quantum mechanics. The other words the Parker scalar product $\langle \varphi, \psi \rangle$ coincides with the scalar product $\langle \varphi, \psi \rangle_\rho$.

$$\langle \varphi, \psi \rangle = \langle \varphi, \psi \rangle_\rho. \quad (72)$$

One of the consequences of the (72) is that the relation $\langle \varphi, (\hat{H}\psi) \rangle = \langle (\hat{H}\varphi), \psi \rangle$ has to be fulfilled in the case when the scalar products are defined in accordance with the rules (66), i.e.

$$\langle \varphi, (\hat{H}\psi) \rangle = \int dx^3 \sqrt{-g} \left( \varphi^+ \gamma_{\underline{0}} \gamma^0 \hat{H}\psi \right),$$
$$\langle (\hat{H}\varphi), \psi \rangle = \int dx^3 \sqrt{-g} \left( \varphi^+ \hat{H}^+ \gamma_{\underline{0}} \gamma^0 \psi \right). \quad (73)$$

### 6.2. HERMITICITY OF HAMILTONIAN CONCERNING PARKER SCALAR PRODUCT

Let us check a satisfiability of the relation $\langle \varphi, (\hat{H}\psi) \rangle = \langle (\hat{H}\varphi), \psi \rangle$ by direct calculation in the case when the scalar product is used in Parker form (66). With this purpose let us consider an expression

$$\Delta \equiv \langle \varphi, (\hat{H}\psi) \rangle - \langle (\hat{H}\varphi), \psi \rangle. \tag{74}$$

If we manage to prove that the expression (74) for $\Delta$ is equal to zero at any functions $\varphi$ and $\psi$, so, Hermiticity of Hamiltonian concerning the Parker scalar product will be proved.

We substitute the expressions for Hamiltonian (24) in (74) and write the expression for $\Delta$ as a sum of three summands:

$$\Delta \equiv \Delta_1 + \Delta_2 + \Delta_3. \tag{75}$$

Here:

$$\Delta_1 = \int dx^3 \sqrt{-g} \left( \varphi^+ \gamma_{\underline{0}} \gamma^0 \left\{ -\frac{im}{\sqrt{(-g^{00})}} \gamma^0 \right\} \psi \right) -$$
$$- \int dx^3 \sqrt{-g} \left( \left( \left\{ -\frac{im}{\sqrt{(-g^{00})}} \gamma^0 \right\} \varphi \right)^+ \gamma_{\underline{0}} \gamma^0 \psi \right) \tag{76}$$

$$\Delta_2 = \int dx^3 \sqrt{-g} \left( \varphi^+ \gamma_{\underline{0}} \gamma^0 \{-i\Phi_0\} \psi \right) - \int dx^3 \sqrt{-g} \left( (\{-i\Phi_0\} \varphi)^+ \gamma_{\underline{0}} \gamma^0 \psi \right). \tag{77}$$

$$\Delta_3 = \int dx^3 \sqrt{-g} \left( \varphi^+ \gamma_{\underline{0}} \gamma^0 \left\{ \frac{i}{\sqrt{(-g^{00})}} \gamma^0 \gamma^k \nabla_k \right\} \psi \right)$$
$$- \int dx^3 \sqrt{-g} \left( \left( \left\{ \frac{i}{\sqrt{(-g^{00})}} \gamma^0 \gamma^k \nabla_k \right\} \varphi \right)^+ \gamma_{\underline{0}} \gamma^0 \psi \right). \tag{78}$$

Let us now calculate each of three terms (76)-(78). For calculating $\Delta_1$ we use the following relations:

$$\gamma^0 \gamma^0 = g^{00}, \quad \gamma_{\underline{0}} \gamma_{\underline{0}} = -E. \tag{79}$$

Thus,

$$\Delta_1 = \int dx^3 \sqrt{-g} \frac{im}{\sqrt{(-g^{00})}} \{-(\varphi^+ \gamma_{\underline{0}} \gamma^0 \gamma^0 \psi) - (\varphi^+ \gamma^{0+} \gamma_{\underline{0}} \gamma^0 \psi)\} =$$
$$= \int dx^3 \sqrt{-g} \frac{im}{\sqrt{(-g^{00})}} \{-g^{00} (\varphi^+ \gamma_{\underline{0}} \psi) + g^{00} (\varphi^+ \gamma_{\underline{0}} \psi)\} \equiv 0. \tag{80}$$

For $\Delta_2$

$$\Delta_2 = i \int dx^3 \sqrt{-g} \{(-\varphi^+ \gamma_{\underline{0}} \gamma^0 \Phi_0 \psi) - (\varphi^+ \Phi_0^+ \gamma_{\underline{0}} \gamma^0 \psi)\} =$$
$$= i \int dx^3 \sqrt{-g} \left( \varphi^+ \gamma_{\underline{0}} [\Phi_0, \gamma^0]_- \psi \right).$$

Using (65), this expression for $\Delta_2$ can be written as

$$\Delta_2 = -i \int dx^3 \sqrt{-g} \left( \varphi^+ \gamma_{\underline{0}} (\gamma^0_{;0}) \psi \right). \tag{81}$$

Let us calculate $\Delta_3$ using the same techniques that have been used to obtain $\Delta_1$ and $\Delta_2$:

$$\Delta_3 = \int dx^3 \sqrt{-g} \, \frac{i}{(-g^{00})} \left\{ \left( \varphi^+ \gamma_{\underline{0}} \gamma^0 \gamma^0 \gamma^k (\nabla_k \psi) \right) + \left( (\nabla_k \varphi^+) \gamma^{k+} \gamma^{0+} \gamma_{\underline{0}} \gamma^0 \psi \right) \right\} =$$

$$= \int dx^3 \sqrt{-g} \, \frac{i}{(-g^{00})} \left\{ \left( \varphi^+ \gamma_{\underline{0}} \gamma^0 \gamma^0 \gamma^k (\nabla_k \psi) \right) + \left( (\nabla_k \varphi^+) \gamma_{\underline{0}} \gamma^k \gamma_{\underline{0}} \gamma_{\underline{0}} \gamma^0 \gamma_{\underline{0}} \gamma^0 \psi \right) \right\} = \qquad (82)$$

$$= \int dx^3 \sqrt{-g} \, \frac{i}{(-g^{00})} \left\{ g^{00} \left( \varphi^+ \gamma_{\underline{0}} \gamma^k (\nabla_k \psi) \right) + g^{00} \left( (\nabla_k \varphi^+) \gamma_{\underline{0}} \gamma^k \psi \right) \right\} =$$

$$= -i \int dx^3 \sqrt{-g} \left\{ \left( \varphi^+ \gamma_{\underline{0}} \gamma^k (\nabla_k \psi) \right) + \left( (\nabla_k \varphi^+) \gamma_{\underline{0}} \gamma^k \psi \right) \right\}.$$

The expression (82) can be written as

$$\Delta_3 = -i \int dx^3 \sqrt{-g} \, j^k{}_{;k} . \qquad (83)$$

Here we used the notation $j^k = \left( \varphi^+ \gamma_{\underline{0}} \gamma^k \psi \right)$ and also

$$\left( \varphi^+ \gamma_{\underline{0}} \gamma^k (\nabla_k \psi) \right) + \left( (\nabla_k \varphi^+) \gamma_{\underline{0}} \gamma^k \psi \right) = \nabla_k j^k = j^k{}_{;k} . \qquad (84)$$

By substituting expressions (80), (81) and (83) in (75), we obtain:

$$\Delta = -i \int dx^3 \sqrt{-g} \left( \varphi^+ \gamma_{\underline{0}} \left( \gamma^0{}_{;0} \right) \psi \right) - i \int dx^3 \sqrt{-g} \, j^k{}_{;k}. \qquad (85)$$

Usually it is assumed that the second summand in the right part of (85) can be reduced to a surface integral according to Gauss theorem and equated to zero. Such an assumption is used particularly in [3]. But Gauss theorem can be used only if an integrand is a divergence of density of covariant vector. If in our case such density is $\left( \sqrt{-g} \, j^k \right)$, Gauss theorem has to be written as

$$\int dx^3 \left( \sqrt{-g} \, j^k \right)_{,k} = \oiint ds_k \left( \sqrt{-g} \, j^k \right). \qquad (86)$$

Let us select a term $-i \int dx^3 \left( \sqrt{-g} \, j^k \right)_{,k}$ in $-i \int dx^3 \sqrt{-g} \, j^k{}_{;k}$.

$$-i \int dx^3 \sqrt{-g} \, j^k{}_{;k} = -i \int dx^3 \sqrt{-g} \left\{ j^k{}_{,k} + \binom{k}{k0} j^0 + \binom{k}{km} j^m \right\} =$$

$$= -i \int dx^3 \sqrt{-g} \left\{ \frac{1}{\sqrt{-g}} \left( \sqrt{-g} \, j^k \right)_{,k} - \frac{\left( \sqrt{-g} \right)_{,k}}{\sqrt{-g}} j^k + \binom{k}{k0} j^0 + \binom{k}{km} j^m \right\} = \qquad (87)$$

$$= -i \int dx^3 \left( \sqrt{-g} \, j^k \right)_{,k} - i \int dx^3 \sqrt{-g} \left\{ -\frac{\left( \sqrt{-g} \right)_{,k}}{\sqrt{-g}} j^k + \binom{k}{k0} j^0 + \binom{k}{km} j^m \right\}.$$

Further we use the following relation

$$\frac{\left( \sqrt{-g} \right)_{,k}}{\sqrt{-g}} = \frac{1}{2} g^{\mu\nu} g_{\mu\nu,k} = \binom{\varepsilon}{\varepsilon k}. \qquad (88)$$

Let us substitute this expression in (87):

$$-i\int dx^3 \sqrt{-g}\, j^k{}_{;k} = -i\int dx^3 \left(\sqrt{-g}\, j^k\right)_{,k}$$
$$-i\int dx^3 \sqrt{-g}\left\{-\begin{pmatrix}\varepsilon\\ \varepsilon k\end{pmatrix} j^k + \begin{pmatrix}k\\ k0\end{pmatrix} j^0 + \begin{pmatrix}k\\ km\end{pmatrix} j^m\right\} = \quad (89)$$
$$= -i\int dx^3 \left(\sqrt{-g}\, j^k\right)_{,k} - i\int dx^3 \sqrt{-g}\left\{-\begin{pmatrix}0\\ 0k\end{pmatrix} j^k + \begin{pmatrix}k\\ k0\end{pmatrix} j^0\right\}.$$

The expression (85) for $\Delta$ has the following view:

$$\Delta = -i\int dx^3 \left(\sqrt{-g}\, j^k\right)_{,k}$$
$$-i\int dx^3 \sqrt{-g}\left(\varphi^+ \gamma_{\underline{0}}\left(\gamma^0{}_{,0} + \begin{pmatrix}0\\ 00\end{pmatrix}\gamma^0 + \begin{pmatrix}0\\ 0k\end{pmatrix}\gamma^k\right)\psi\right) \quad (90)$$
$$-i\int dx^3 \sqrt{-g}\left\{-\begin{pmatrix}0\\ 0k\end{pmatrix} j^k + \begin{pmatrix}k\\ k0\end{pmatrix} j^0\right\}.$$

For the Kerr solution in considered approximation

$$\gamma^0{}_{,0} = 0, \quad \begin{pmatrix}0\\ 00\end{pmatrix} = 0, \quad \begin{pmatrix}k\\ k0\end{pmatrix} = 0. \quad (91)$$

Using (91) we obtain:

$$\Delta = -i\int dx^3 \left(\sqrt{-g}\, j^k\right)_{,k}$$
$$-i\int dx^3 \sqrt{-g}\begin{pmatrix}0\\ 0k\end{pmatrix} j^k - i\int dx^3 \sqrt{-g}\left\{-\begin{pmatrix}0\\ 0k\end{pmatrix} j^k\right\} = 0. \quad (92)$$

Here we used the Gauss theorem in form (86) and assume the surface integral is equal to zero.

The relation (92) means that we proved Hermiticity of initial Hamiltonian (24) concerning the Parker scalar product.

## 7. DISCUSSION

The results obtained in this study make it possible to look at the description of quantum mechanics of spin 1/2 particles in stationary gravitational fields from a new point of view.

For example, before this study, the problem of non-uniqueness of Hamiltonians and their sensitivity to the choice of tetrad vectors in our opinion was unresolved. The apparatus of pseudo-Hermitian quantum mechanics used in this study allowed us to resolve this issue at least as applied to the Schwarzschild and the Kerr solutions. It was proven that the resulting Hamiltonian $\hat{H}$ defined as (58) does not depend on the choice of tetrad vectors and is Hermitian. The scalar products also do not dependent on the choice of tetrad vectors, if they were calculated using operators $\rho$ shown in Table 1.

In our opinion, the uniqueness of the Hamiltonian $\hat{H}$ is nontrivial. Indeed, all of the three initial expressions for Hamiltonians (47), (50), (52), first, differ from each other and, second, are non-Hermitian concerning standard scalar product in Hilbert space. After applying procedures for the transformion of initial Hamiltonians $\hat{H}$ into their Hermitian expressions $\hat{H}$, the latter could in principle differ in some Hermitian summands. It is not the case, however, and the expressions for $\hat{H}$ are the same in all the three cases. Such a coincidence means that whatever the choice of tetrad vectors in a gravitational field there will always exist a single Hermitian Hamiltonian $\hat{H}$, which has the same spectrum of energy levels as any of the starting operators $\hat{H}$.

Upon transition to the Hamiltonian $\hat{H}$, one can use the quantum mechanics apparatus in its standard form. In particular, the left member of the Schrödinger equation will contain the operator $i(\partial/\partial t)$, in which the time coordinate $t$ is understood to be the time of an infinitely distant observer.

Interestingly, the expression derived in this study for the Hamiltonian $\hat{H}$ is the same as the expression proposed in [9] for the Kerr field[7]. The formalism of pseudo-Hermitian Hamiltonians in fact validates the expressions for Hermitian Hamiltonians used in [4], [9].

The comparison of quantum mechanics treatment of dynamic of Dirac particle in stationary gravitational field is presented in Table 2. The approaches in the columns A, B are equivalent in the sense that they result in the same spectrum of energy of Hamiltonians and the same values of scalar products. We proved that the approaches in columns B, C are also equivalent, because the operator $\rho$ in scalar product (9) obtained within the formalism of pseudo-Hermitian quantum mechanics coincides with the operator $\sqrt{-g}\gamma_0\gamma^0$. The expression for this operator follows from the scalar product introduced in papers [2], [3]. We suppose that this result is also very important because the construction algorithm of the Parker scalar product satisfies a number of general-theoretical requirements (see. Section 6.1), which makes the use of this algorithm attractive.

Table 2 – Comparison of quantum mechanics treatment of dynamics of Dirac particle in stationary gravitational field

|  | Methods of quantum mechanics treatment | | |
|---|---|---|---|
|  | A | B | C |
|  | Hermitian quantum mechanics in $\eta$ - representation | Pseudo-Hermitian quantum mechanics in initial representation | Approach based on Parker scalar product |
| Hamiltonian | $\hat{H}=\hat{H}^+$ | $\hat{H}=\rho^{-1}\hat{H}^+\rho$ | $\hat{H}=\left(\sqrt{-g}\gamma_0\gamma^0\right)^{-1}\hat{H}\left(\sqrt{-g}\gamma_0\gamma^0\right)$ |
| Dependence of Hamiltonian type on choice of system of tetrad vectors | No | Yes | Yes |

---

[7] The statement concerning the Kerr field also holds for the Schwarzschild field, for which a Hermitian Hamiltonian was proposed in [4].

| Scalar product | Standard scalar product for Hilbert space: $$(\Phi,\Psi) = \int d^3x \left(\Phi^+\Psi\right)$$ | With an weight operator $\rho = \eta^+\eta$: $$\langle\varphi,\psi\rangle_\rho = \int d^3x \left(\varphi^+\rho\psi\right)$$ | $\rho = \eta^+\eta = \left(\sqrt{-g}\gamma_0\gamma^0\right)$ $$\langle\varphi,\psi\rangle = \int d^3x \sqrt{-g}\left(\varphi^+\gamma_0\gamma^0\psi\right)$$ |
|---|---|---|---|
| Connection between Hamiltonians | $\hat{H} = \eta\hat{H}\eta^{-1}$ | | |
| Connection between scalar products | $(\Phi,\Psi) = \langle\varphi,\psi\rangle_\rho = \langle\varphi,\psi\rangle$ | | |
| Connection between wave functions | $\Psi = \eta\psi$ | | |

The results of this study allow us to claim that the method of pseudo-Hermitian Hamiltonians enables the application of the relativistic quantum mechanics formalism practically in its standard form. The expression for the operator $\hat{H}$ in $\eta$ - representation allows to get rid of "ambiguity" connected with different type of the initial Hamiltonians at the use of different system of tetrad vectors. We think that this feature of the pseudo-Hermitian method makes it preferable as applied to the problems, in which gravitational effects are resolved and their quantitative characteristics are analyzed.